\documentclass[aps,prl,twocolumn,showpacs,preprintnumbers,amsmath,amssymb]{revtex4}
\usepackage{graphicx}% Include figure files
\usepackage{dcolumn}% Align table columns on decimal point
\usepackage{bm}% bold math

\begin{document}

%\preprint{APS/123-QED}

\title{DMRG-study of current and activity fluctuations near non-equilibrium phase transitions}% Force line breaks with \\

\author{Mieke Gorissen$^{1}$}
% \altaffiliation[Also at ]{Physics Department, XYZ University.}
\author{Jef Hooyberghs$^{2,1,3}$}%
\author{Carlo Vanderzande$^{1,3}$}
\affiliation{%
$^1$ Departement WNI, Universiteit Hasselt, 3590 Diepenbeek, Belgium.\\
$^2$ VITO, Boeretang 200, 2400 Mol, Belgium. \\
$^3$ Instituut Theoretische Fysica, K.U. Leuven, 3001 Heverlee, Belgium.
} 
%\\

%\author{Carlo Vanderzande}
%\affiliation{Departement WNI, Hasselt University, 3590 Diepenbeek, Belgium 
%\\
%Instituut Theoretische Fysica, Katholieke Universiteit Leuven, 3001 Heverlee, Belgium
%}%

\date{\today}% It is always \today, today,

\begin{abstract}
Cumulants of a fluctuating current  can be obtained from a free energy-like generating function which for Markov processes equals the largest eigenvalue of a generalized generator. We determine this eigenvalue with the DMRG for stochastic systems.  We calculate the variance of the current in the different phases, and at the phase transitions, of the totally asymmetric exclusion process.  Our results can be described in the terms of a scaling ansatz that involves the dynamical exponent $z$. We also calculate the generating function of the activity near the absorbing state transition of the contact process. Its scaling properties can be expressed in terms of known critical exponents. \end{abstract}

\maketitle
Physical systems that are in contact with two reservoirs at a different temperature or chemical potential, develop a heat or particle current \cite{deGroot1984}. In macroscopic systems, fluctuations of these currents can often be neglected. As is the case in equilibrium systems, one can however expect that such fluctuations become important in mesoscopic systems and in the vicinity of a non-equilibrium critical point \cite{Marro1}. 

The statistics of current fluctuations in mesoscopic conductors have received a lot of attention in the past decade \cite{Nazarov1}, since they can, for example, give insight on correlated electron transport.
It is nowadays possible to measure experimentally third and higher order cumulants of the current in problems of charge transport \cite{Heikkila1,Fujisawa1}. 
%These cumulants could play a similar role outside equilibrium as specific heats or susceptibilities do in %equilibrium.
Theoretically, these cumulants can be obtained as derivatives of a generating function. This function has many similarities to the free energy in equilibrium systems.

In the present Letter, we focus on the scaling of the current distribution in one-dimensional (classical) stochastic models such as the (a)symmetric exclusion process. This stochastic process
%was originally introduced as a model for biopolymerisation \cite{MacDonald1}, but 
is a standard model of  non-equilibrium statistical mechanics \cite{Derrida2,Blythe1}. Rigorous results are known for the current distribution in this model both on a ring and for open boundaries \cite{Derrida1,Appert1,Bodineau1,Derrida3}. Moreover, several approximate and numerical approaches to this problem have been developped: simulation techniques that sample rare events \cite{Giardina1,Lecomte1}, renormalisation approaches \cite{Lecomte2} and perturbation techniques \cite{ Baiesi1}. Here we apply for the first time the density matrix renormalisation group (DMRG) to the investigation of current fluctuations. We illustrate the method for the current of the totally asymmetric exclusion process, but the technique is more general. As an example we also present results on the total number of changes of configuration (a quantity that has been called activity \cite{Lecomte3}) in the contact process \cite{Marro1}.

In the totally asymmetric exclusion process (TASEP), each site of a one-dimensional lattice of $L$ sites can be empty or occupied by at most one particle. The dynamics of the model is a continuous time Markov process in which a particle hops to its right neighbor with unit rate 
%(this fixes the time scale) 
provided that site is empty. At the left boundary particles enter the system with rate $\alpha$, while at the right boundary they leave it with rate $\beta$. 
Asymptotically, the TASEP reaches a non-equilibrium steady state (NESS) in which a nonzero current flows through the system.  
It is by now well known that the TASEP has three distinct phases \cite{Derrida4}: in the low density (LD) phase ($\alpha<1/2, \beta>\alpha$) the average current $J$ (per bond and in the thermodynamic limit) equals $\alpha(1-\alpha)$ while in the high density (HD) phase ($\beta<1/2, \alpha>\beta$) it is $\beta(1-\beta)$. Finally, in the maximal current (MC) phase, $J=1/4$. 

Let  $J_L(t)$ be the total current through all bonds up to time $t$ during a  realisation of the stochastic process. The statistical properties of this current 
%in the NESS 
can be obtained from its generating function
\begin{eqnarray}
\mu(s,L) \equiv \lim_{t \to \infty} \frac{1}{t} \log \langle e^{sJ_L(t)}\rangle
\label{0}
\end{eqnarray}
where the average is taken over the realisations of the stochastic process. The distribution of $J_L(t)$ at large times can be determined from $\mu(s,L)$ by a Legendre transformation while
the average current $J(L)$, its variance $\Delta(L)$ and higher cumulants can be found as derivatives of $\mu(s,L)$:
\begin{eqnarray*}
J(L)&=&\lim_{t \to \infty} \frac{1}{t}\langle J_L(t) \rangle=\frac{\partial \mu}{\partial s} (0,L)
\end{eqnarray*}
\begin{eqnarray}
\Delta(L) &=& \lim_{t \to \infty} \frac{1}{t} \left(\langle J_L^2(t)\rangle - \langle J_L(t)\rangle^2\right)=\frac{\partial^2 \mu}{\partial s^2}(0,L)
\label{11}
\end{eqnarray}

Let  $\sigma_i$ be a spin variable which equals  $1$ if the site $i$ is vacant and $-1$ if it is occupied. The state of the system is then characterized by the probability $P({\cal C};t)$ to be in a given microstate ${\cal C}=\{\sigma_1,\ldots,\sigma_L\}$. 
This probability evolves according to the master equation \cite{VanKampen1}
\begin{eqnarray}
\frac{d P({\cal C};t)}{d t} = H P({\cal C};t)
\end{eqnarray}
where $H$ is the generator of the stochastic process. The properties of the NESS of the stochastic process can be determined from the (right) eigenvector of $H$ with the largest eigenvalue \cite{VanKampen1}. Similarly,
it is not difficult to show that generating functions such as $\mu(s,L)$ 
%(or that of the activity)
can be obtained as the largest eigenvalue of a modified generator $H(s)$ \cite{Derrida2}. For the current $J_L(t)$, $H(s)$  equals
\begin{eqnarray}
H(s)&=& \sum_{i=1}^{L-1} \left[ e^ss_i^+ s_{i+1}^-- n_i v_{i+1}\right] + \alpha\left[e^s s_1^--n_1\right] \nonumber \\  &+ &\beta\left
[e^ss_L^--v_L\right]
\label{3}
\end{eqnarray}
Here we have used the "quantum" notation for stochastic systems \cite{Schutz1}. The operators $s_i^+$ and $s_i^-$ respectively destroy and create a particle at site $i$, while $n_i$ and $v_i$ count the number of particles and vacancies at that site.

Formulated this way, determining $\mu(s,L)$ is mathematically similar to finding the ground state energy of a quantum spin or fermion chain.
One of the most succesful numerical techniques to study low temperature properties of quantum chains is the DMRG \cite{White1,Schollwock}. More recently, this method has been extended to stochastic systems \cite{Carlon1,Kaulke1} where the main difference is that in general the generator $H$ is non-Hermitian. Here we apply the method for the first time to operators such as $H(s)$ which are neither Hermitian nor stochastic. We found that with the DMRG it is possible to obtain $\mu(s,L)$ numerically exact for systems up to $L=60$ with only modest computing facilities. Since there are no essential new ingredients in the method as such \cite{Gorissen1}, we focus here on the results.

Firstly, in order to test the method we have calculated $\mu(s,L)$ for the symmetric exclusion process (SEP) for which this function is known for large $L$ values \cite{Derrida5}. In the SEP, particles can hop both to the right and left with equal rate. At its boundaries, the system is in contact with particle baths of density $\rho_a$ and $\rho_b$. In Fig. 1, we show typical results for $L\mu(s,L)$ for various $L$-values, together with the exact result (full line). As can be seen, there is a fast convergence towards the asymptotic results.
\begin{figure}[here]
\includegraphics[angle=-90,width=0.45\textwidth]{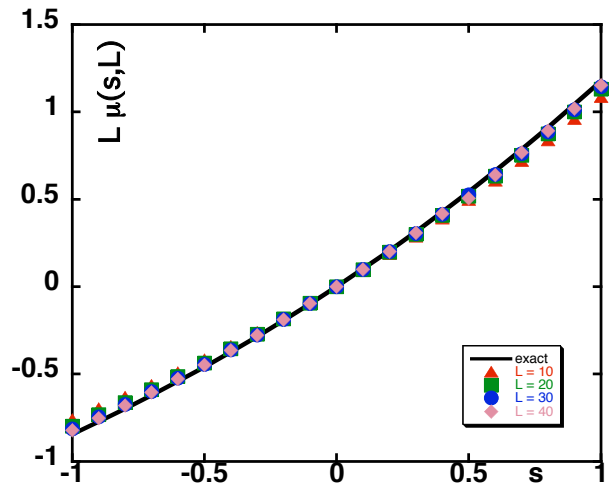}
\caption{\label{fig1} Cumulant generating function for the symmetric exclusion process with $\rho_a=1, \rho_b=0$. 
Shown are the asymptotic results of \cite{Derrida5} (full line) and DMRG results for different system sizes.
}
\end{figure}

Going back to the TASEP, we calculated $\mu(s,L)$ in the various regions of the phase diagram. The cumulants $J(L)$ and $\Delta(L)$ are then determined by numerical differentiation. 
\begin{figure}[here]
\includegraphics[angle=-90,width=0.45\textwidth]{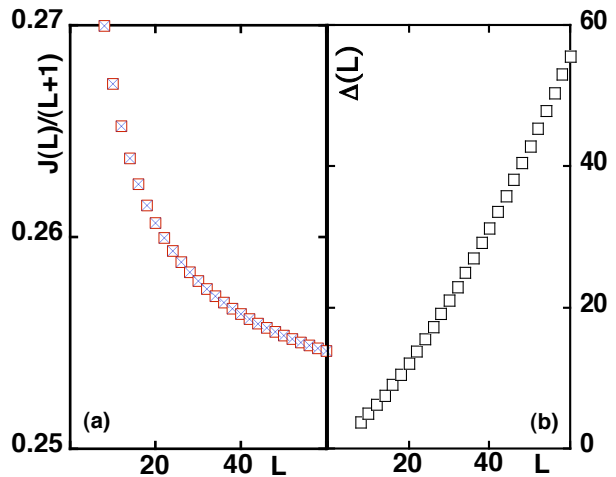}
\caption{\label{fig1} (a) Average current per bond from DMRG (squares) compared with exact results from \cite{Derrida4} (crosses). (b) Variance of the total current. Both results are for $\alpha=3/5$ and $\beta=2/3$. }
\end{figure}
As an example, we present in Fig. 2 our results for  $J(L)/(L+1)$  and $\Delta(L)$ in the MC phase. Also shown are the exact results for $J(L)$ per bond obtained from the matrix product ansatz \cite{Derrida4}. The numerical data coincide with the exact ones within the accuracy. 
The variance of the current was so far not determined exactly. We find that $\Delta(L)$ increases as $L^\sigma$. The corrections to this power law behavior are strong and cannot be neglected for the system sizes we studied. In order to get reliable exponent estimates we have used the BST-algorithm \cite{Henkel1}. We find that in the MC phase, but also at the transition line between the MC and LD (or HD)-phase,  $\sigma=1.50(2)$. In the LD (and HD) phase,  $\sigma$ changes to $2.01(4)$. Finally, along the coexistence line between HD and LD phases, 
%(i.e. the line $\alpha=\beta<1/2$)
we find $\sigma=2.03(3)$. 
These results strongly suggest that $M=\lim_{L \to \infty} \Delta(L)/L^2$ behaves as an order parameter: it is zero in the MC-phase and non-zero in both the LD and HD phases.

Given the similarities between the generating function and the free energy, it is natural to ask about the scaling properties of $\mu$ near a phase transition. To focus attention, we consider the transition line between LD and MC phases ($\alpha=1/2, \beta>1/2$). We propose that under a rescaling with a factor $b$ the singular part of $\mu$  transforms as
\begin{eqnarray}
\mu(s,\Delta \alpha,L) \sim b^{-z} \mu(b^{y_s}s, b^{y_\alpha} \Delta \alpha,L/b)
\label{10}
\end{eqnarray}
where $\Delta\alpha=\alpha-1/2$ and $z$ is the dynamical exponent. We conjecture that $z$ replaces the dimension $d$ that appears in the scaling of the free energy because $\mu(s,L)$ is a quantity per "unit of time", whereas the free energy is per unit of volume. The exponent $y_s$ is a new exponent associated with current fluctuations, and $y_\alpha$ is like a thermal exponent in equilibrium critical phenomena. 
From (\ref{11}) and (\ref{10}), it follows that $J(L) \sim L^{-z+y_s}$ and $\Delta(L) \sim L^{-z + 2 y_s}$ at the transition. From the exact results on $J(L)$ and our data on the variance, we find $z=1.50(2)$ and $y_s=1.50(2)$. This value of $z$ agrees with that determined by the Bethe-ansatz \cite{deGier1}, $z =3/2$, thus providing strong support to the scaling form (\ref{10}). We conjecture that also $y_s=3/2$.  
%According to our scaling picture $\partial J(L)/\partial \alpha \sim L^{-z+y_s+y_\alpha}$ at the transition line. 
Finally, $y_\alpha$ can be obtained from 
$\partial J(L)/\partial \alpha$. This derivative can easily be calculated from the exact results, and gives $y_\alpha=1/2$
\cite{Gorissen1}.

To test our scaling ansatz further, we investigate the variance of the current which should scale as 
\begin{eqnarray*}
\Delta(L,\Delta \alpha) \sim L^{-z+2y_s} H(L^{y_\alpha}\Delta \alpha)
\end{eqnarray*}
%Consistency with the numerical results then implies that for large positive arguments, the scaling function $H(x)$ becomes constant, whereas $H(x) \sim x^{1/(2y_a)}$ for  $x \to -\infty$.
with $H$ a scaling function. To match the numerically determined behavior of the variance in the different phases, $H(x)$ should be constant for $x>0$, and  linear in $x$ for small $x<0$. This implies that  $M$ goes to zero linearly as the LD-MC transition line is approached from below.
In Fig. 3, we show our data for $\Delta(L,\Delta \alpha) L^{z-2y_s}$ versus $L^{y_\alpha }$ ($\Delta \alpha <0$).
%at $s=0$, and in Fig. 4  as a function of $s$ at $\Delta \alpha=0$. 
The scaling is well satisfied and gives exponent values close to those determined above
\cite{Opmerking}.
\begin{figure}[here]
\includegraphics[angle=-90,width=0.45\textwidth]{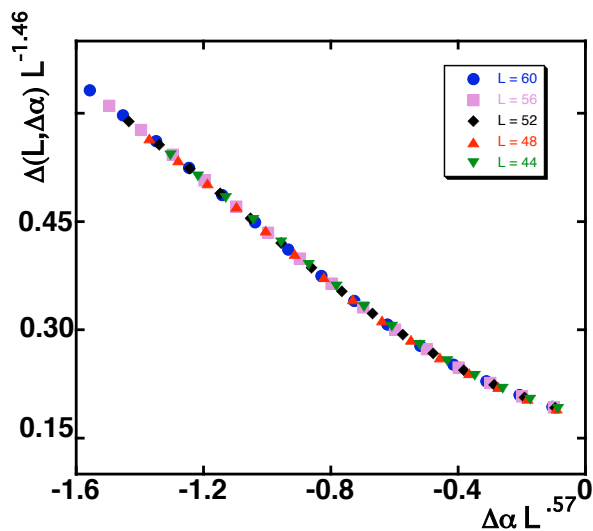}
\caption{\label{fig3} Scaling plot of $\Delta(L,\Delta\alpha) L^{z-2 y_s}$ versus $L^{y_\alpha}\Delta \alpha$ at $s=0, \Delta\alpha<0, \beta=2/3$.}
\end{figure}
In a similar way, we also checked that the scaling of $\partial^2 \mu/\partial s^2$ as a function of $s$ at the transition line is well satisfied \cite{Gorissen1}. Finally, 
also data on the third cumulant of the current can be well described by our scaling proposal \cite{Gorissen1}.
%\begin{figure}[here]
%\includegraphics[width=8.0cm,height=6.0cm]{Figure3}
%\caption{\label{fig4} Scaling plot of $\Delta(s,L) L^{-z+2 y_s}$ versus $sL^{y_s}$ at $\alpha=0.5, 
%\beta=2/3$.}\end{figure}

As a second application of our approach we study
the one-dimensional contact process (CP). In this model,  each site of a lattice can be occupied by at most one particle. An occupied site becomes empty with rate $1$, while an empty site becomes occupied with a rate $\zeta \lambda/2$. Here $\zeta$ is the number of occupied neighbors. When $\lambda<\lambda_c$ the process reaches an absorbing state in which all sites are empty. For $\lambda > \lambda_c$, and in an infinite system, the model reaches a NESS with a finite density $\rho$ of particles. The contact process \cite{Harris2} is a standard model for phase transitions out of an absorbing state \cite{Marro1}. It is known from extensive numerical investigations that its phase transition belongs to the universality class of directed percolation \cite{Grassberger1}. The scaling properties of various quantities near $\lambda_c$ are well characterized \cite{Marro1}. 
Here we are interested in the activity $K_L(t)$ of the model, which gives the total number of changes of configuration in a realization of the process up to time $t$. The generating function of $K_L(t)$ is 
\begin{eqnarray}
\pi(s,L) \equiv \lim_{t \to \infty} \frac{1}{t} \log \langle e^{sK_L(t)}\rangle
\nonumber
\end{eqnarray}
This function can again be obtained as the largest eigenvalue of a generator which in this case equals
\begin{eqnarray}
\sum_{i=1}^{L} \left[(e^s s_i^+-n_i)-\frac{\lambda}{2}(n_{i-1}+n_{i+1})(e^s s_i^- - v_i)\right]
\end{eqnarray}
($n_0=n_{L+1}=0$). A finite system will always reach the absorbing state asymptotically. To avoid this, we allow the creation of particles at the boundary sites. Following the reasonings made for the TASEP, we expect that near the absorbing state transition, $\pi$ scales as
\begin{eqnarray}
\pi(s,\Delta \lambda,L) = b^{-z} \pi(b^{y_K} s, b^{1/\nu_\perp} \Delta \lambda,L/b)
\label{5}
\end{eqnarray}
Here $\Delta \lambda=\lambda-\lambda_c$. The exponents $z=\nu_\parallel/\nu_\perp=1.5805$ and $\nu_\perp=1.09684(6)$ are known numerically \cite{Jensen2} while $y_K$ is a new exponent.

It is possible to express $y_K$ in terms of other, known, exponents. From the dynamics of the model one can show that  $\langle K_L(t)\rangle$ \cite{Gorissen1} obeys
\begin{eqnarray}
\frac{d\langle K_L \rangle}{dt}= 2 \sum_{i=1}^L \langle n_i\rangle + \sum_{i=1}^L \frac{d\langle n_i\rangle}{dt}
\label{4}
\end{eqnarray}
In the NESS, the second term in (\ref{4}) approaches zero, whereas the first one becomes equal to $2L \rho$. The scaling of $\rho$ is well known and therefore the average activity should scale as
\begin{eqnarray}
K(L)=\lim_{tÊ\to \infty} \frac{\langle K_L(t)\rangle}{t} = L^{1-\beta/\nu_\perp} F(L^{1/\nu_\perp} \Delta \lambda)
\label{6}
\end{eqnarray}
Here $F$ is a scaling function and $\beta=0.27649$ \cite{Jensen2}. Since $K(L)$ is also the first derivative of $\pi$ we get from (\ref{5}) and (\ref{6}): $y_K=1+(\nu_\parallel-\beta)/\nu_\perp=2.3284$. 
%The variance of $K_L(t)$ then should scale as
%\begin{eqnarray*}
%\Delta_K(L)=\lim_{t \to \infty} \frac{\langle K_L^2(t)\rangle-\langle K_L(t)\rangle^2}{t}=L^{2+\frac{\nu_
%\parallel - 2 \beta}{\nu_\perp}}
%\end{eqnarray*}
%at the critical point $s=\Delta \lambda=0$. 
We have calculated $\pi(s,\Delta \lambda,L)$ using the DMRG. 
In Fig. 4, we show our results for the variance of the activity as a function of $\Delta \lambda$ and $L$. 
\begin{figure}[here]
\includegraphics[angle=-90,width=0.45\textwidth]{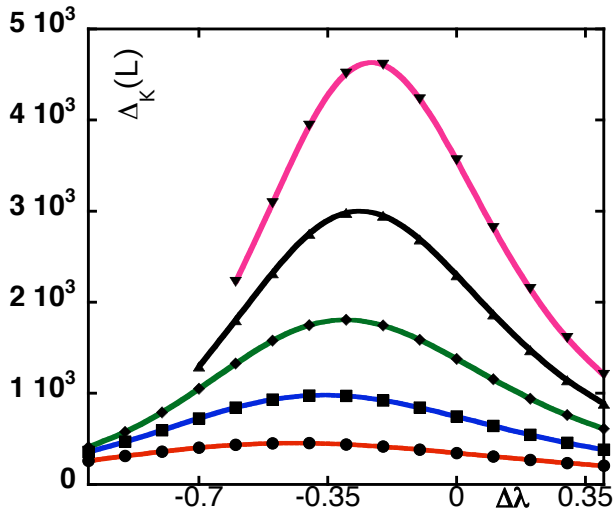}
\caption{\label{fig5} Plot of the variance $\Delta_K(L)$ of the activity of the contact proces for (top to bottom) $L=44,38,32,26,20$.}
\end{figure}
At criticality, we find that the average activity diverges as $L^{.746(2)}$, while its variance goes as $L^{3.08(2)}$. These exponents are close to $-z+y_K=0.7479$ and $-z+2y_K=3.0763$ predicted by the scaling (\ref{5}).
Other evidence of (\ref{5}) can be seen in Fig. 5 where we present a scaling plot of $K(s,L)=\partial \pi/\partial s$ as a function of $s$ at $\Delta \lambda=0$. This quantity should scale as $L^{-z+y_K} G(L^{y_K} s)$. The numerical data again support this prediction \cite{Opmerking}.
%Using the values of $y_K$ and $z$ determined above, the predicted scaling is found to be well 
%satisfied. We therefore conclude that the activity fluctuations in the contact process are very well 
%described by the ansatz  (\ref{5}) and that the exponents appearing in that equation can be determined %completely in terms of (numerically) known exponents of the CP.

\begin{figure}[here]
\includegraphics[angle=-90,width=0.45\textwidth]{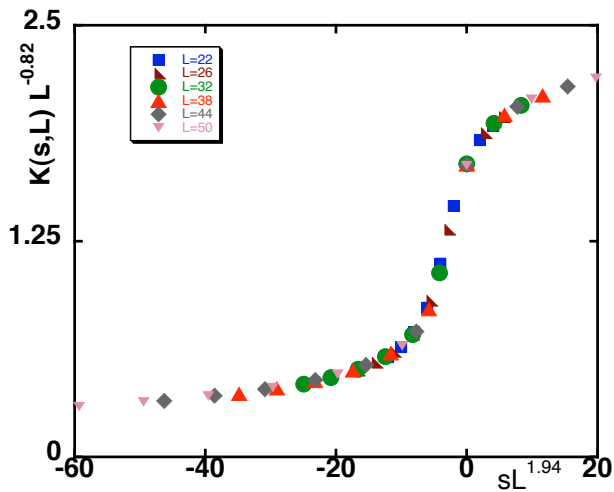}
\caption{\label{fig4} Scaling plot of $\partial \pi/\partial s (s,\Delta \lambda=0,L)L^{z-y_K}$ versus $sL^ {y_K}$ for $22<L<50$.
}
\end{figure}

In summary, we determined the generating function of the current in the TASEP and of the activity in the CP with the DMRG. We  proposed a scaling form for these generating functions which is supported by all numerical data. We believe that this scaling is quite general and can be applied to other models as well. 
%It would, for example, be interesting to extend the scaling ansatz to models with two or more currents. 
%It should also be possible to get insight in the time dependent behavior of current cumulants from an 
%investigations of the gaps in the spectrum of operators such as $H(s)$. Finally, a more fundamental 
%understanding for the scalings proposed here from a renormalisation approach is desirable.

{\bf Acknowledgement} We thank M. Baiesi, E. Carlon, J. de Gier and C. Maes for useful discussions. We thank the 'Expertise centre for Digital Media' (UHasselt) for the use of computer facilities.

%The DMRG is an iterative procedure in which by a suitable truncation of the vector space of a system %of $L$ sites, (part of) the spectrum of a system of $L+2$ sites can be determined. 
%Suppose that the largest eigenvalue of (\ref{3}) together with its right  ($|0_r\rangle$) and left 
%eigenvector ($\langle 0_l|)$ can be determined for a system of $L$ (where $L$ is even) sites. The %'density matrix' of this state is chosen in a symmetric way as 
%\begin{eqnarray*}
%\rho = \frac{1}{2} \left[ |0_r\rangle\langle 0_r| +  |0_l\rangle\langle 0_l|\right]
%\end{eqnarray*}
%From $\rho$, the reduced density matrices $\rho^r$ and $\rho^l$ on the right and left half of the system %are constructed. In the truncation step, $H(s)$ is projected on the $m^2$ dimensional vector space 
%spanned by the direct products of the $m$ eigenvectors of $\rho^l$ and $\rho^r$ with the largest 
%eigenvalues. 
%Finally, two new sites are added in the middle of the system, a $4m^2$-dimensional representation of %$H(s)$ for the resulting system of $L+2$ sites is determined, and its largest eigenvalue is calculated. %The procedure is then iterated for increasing $L$. Because of the consecutives truncations of the 
%vector space it is possible to determine $\mu(s,L)$ accurately for rather large $L$-values \cite
%{Opmerking1}. We used typically $m=16$ or $m=32$ and could easily obtain results for $L \leq 60$ 
%using only a  laptop.

%\bibliography{apssamp}

\begin{thebibliography}{100}
\bibitem{Opmerking}
Because of strong scaling corrections, the exponents that give the best scaling collapse differ slightly from those obtained after BST-extrapolation. The latter are more asymptotic and therefore more 
reliable.
\end{thebibliography}
%\item{DoiEdwards86} Doi M. and Edwards S.F., {\it The theory of polymer dynamics}, Oxford University Press (1986).
% Produces the bibliography via BibTeX.

\end{document}